\pdfoutput=1
\documentclass[aps,prl,twocolumn,superscriptaddress]{revtex4}
\usepackage{color}

\usepackage{graphicx}
\usepackage{boxedminipage}

\begin{document}

\title{One-dimensional half-metallic interfaces of two-dimensional honeycomb insulators}

\author{N. C. Bristowe}
\affiliation{Theory of Condensed Matter,
             Cavendish Laboratory, University of Cambridge, 
             CB3 0HE, UK}
\affiliation{D\'epartment de Physique, Universit\'e de Li\`ege,
             B-4000 Sart-Tilman, Belgium}
\author{Massimiliano Stengel}
\affiliation{ICREA - Instituci\'o Catalana de Recerca i Estudis Avan\c{c}ats, 08010 Barcelona, Spain}
\affiliation{Institut de Ci\`{e}ncia de Materials de Barcelona (ICMAB-CSIC), Campus UAB, 08193 Bellaterra, Spain}
\author{P. B. Littlewood}
\affiliation{Theory of Condensed Matter,
             Cavendish Laboratory, University of Cambridge, 
             CB3 0HE, UK}         
\affiliation{Physical Sciences and Engineering,
             Argonne National Laboratory, 
             Argonne, Illinois 60439, USA}    
\author{Emilio Artacho}
\affiliation{Theory of Condensed Matter,
             Cavendish Laboratory, University of Cambridge, 
             CB3 0HE, UK}
\affiliation{CIC Nanogune, and DIPC,
		Tolosa Hiribidea 76, 20018 San Sebastian, Spain}      
\affiliation{Basque Foundation for Science Ikerbasque, 48011 Bilbao, Spain}
\author{J. M. Pruneda}
\affiliation{Centre d'Investigaci\'{o}n en Nanoci\'{e}ncia i Nanotecnologia (CSIC-ICN), Campus UAB, 08193 Bellaterra, Spain}             

\date{\today}
\begin{abstract}
We study zigzag interfaces between insulating compounds that are isostructural to graphene,
specifically II-VI, III-V and IV-IV two-dimensional (2D) honeycomb insulators.
We show that these one-dimensional interfaces are polar, with a net density of excess charge that
can be simply determined by using the ideal (integer) formal valence charges, regardless of the 
predominant covalent character of the bonding in these materials.
We justify this finding on fundamental physical grounds, by analyzing the topology of the 
formal polarization lattice in the parent bulk materials.
First principles calculations elucidate an electronic compensation mechanism not dissimilar 
to oxide interfaces, which is triggered by a Zener-like charge transfer between interfaces 
of opposite polarity.
In particular, we predict the emergence of one dimensional electron and hole
gases (1DEG), which in some cases are ferromagnetic half-metallic.
\end{abstract}
\maketitle 

Since the first two-dimensional material (graphene) was successfuly synthesized by mechanical 
exfoliation~\cite{Novoselov2004}, a growing number of studies has been devoted to other planar systems. 
These include BN monolayers~\cite{Pacile2008}, layered transition metal oxides~\cite{Osada2009}, 
dichalcogenides~\cite{Lee2010,Coleman2011}, or topological insulators such as Bi$_2$Te$_3$ or 
Bi$_2$Se$_3$~\cite{Teweldebrhan2010}.
Of particular note in this context are a number of II-VI, III-V and IV-IV  
compounds, whose stable bulk phase is wurtzite, but may adopt a planar
graphitic phase in ultrathin films. These were first predicted theoretically
several years ago~\cite{Goniakowski2004,Freeman2006},
and later grown in the lab~\cite{Tusche2007}.

Recent improvements in growth techniques, allow joining these nanosheets together.
In addition to the many heterostructures studied so far involving graphene and BN, 
obtained by modified stacking~\cite{Dean2010} or segregated nanosheets~\cite{Ci2010,Levendorf2012}, 
there are and will be coplanar heterostructures~\cite{Canas-Ventura2007,cSahin2009}, made from different 
insulators on the same sheet many of them having a significant degree of ionic character.
This brings up the important question of whether ``polar discontinuities''~\cite{Ohtomo2004} 
might play a role in these low-dimensional systems. 
Vertical stacking of nanosheets is not dissimilar to oxide interfaces~\cite{Mannhart2010},
and polarity effects will resemble those of ultrathin films already discussed 
in the literature~\cite{Goniakowski2004,Freeman2006}. However, to our knowledge, 
coplanar interfaces between 2D insulators have not been addressed from first principles,
except for a qualitative discussion of the electrostatics~\cite{Goniakowski:2011}.

Compared with the case of polar oxide interfaces, there are a few additional challenges
that are specific to the 2D case. 
First, the electrostatics is somewhat more complicated:
unlike a \emph{plane} of charge (interface between 3D materials), which generates a 
uniform electric field in the whole space (essentially a 1D problem), a \emph{line} of charge 
(interface between 2D materials) produces a logarithmically divergent potential with 
inhomogeneous stray fields in the vacuum region. 
Second, the symmetry group (with a main six-fold rotation axis but no center of symmetry) 
gives rise to different polarization classes than the ones found in centrosymmetric 
insulators~\cite{Banerjee2012}.
%
%
This calls for special care in the application of the modern theory of polarization, 
and in particular of the \emph{interface theorem}~\cite{Vanderbilt1993}, 
for the determination of the excess ``bound'' charge at a given interface.
Moreover, the lack of center of symmetry implies that these materials are 
piezoelectrically active.
%

Here we perform extensive density-functional theory calculations to show that 
(i) there indeed exists a net charge at the 1D zigzag interface between two 
chemically distinct polar insulating honeycomb domains; that (ii) the 
linear charge density is uniquely determined from bulk properties of the 
participating materials, and is related to the nontrivial topology of their respective 
polarization lattice; that (iii) the divergent electrostatic potential 
produced by the lines of charge triggers a Zener-like breakdown mechanism 
in the limit of increasingly wide domains; and that (iv) such mechanism gives rise to 
spin-polarized {\it one-dimensional electron/hole gases} (1DEG/1DHG). 
In what follows, we proceed addressing points (i) to (iv) in the same sequential order.

\begin{figure}[b]
\begin{center}
\includegraphics[width=3.2in]{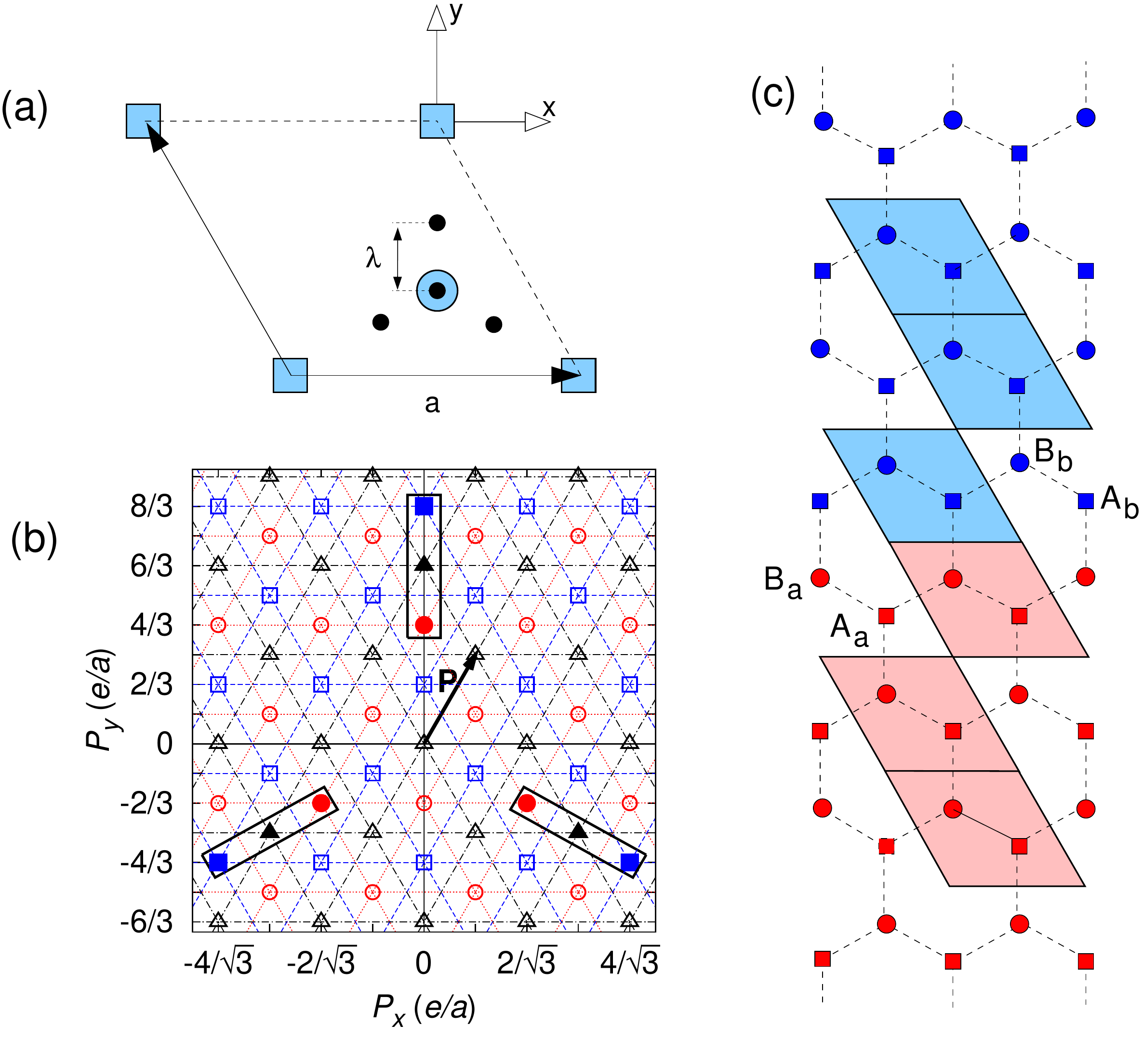}
\caption{\label{honeycomb}{(Color online)
(a) The position of the Wannier centers (small circles) in the primitive unit cell of ZnO, AlN and SiC.
The ``cation'' (Zn, Al, Si) is represented by squares, and the ``anion'' (O, N, C) by circles.  
``a'' is the cell parameter (3.316 \AA, 3.121 \AA, 3.110 \AA\ for ZnO, AlN and SiC respectively),
and $\lambda$ is the distance of the $sp^2$ Wannier centers from the anion 
($\lambda$=0.431 \AA, 0.491 \AA, 0.677 \AA\ respectively), which increases with the bond covalency.
(b) The triangular lattice of allowed polarization values for the II-VI (red circles), III-V (black triangles) 
and IV-IV (blue squares) 2D honeycomb compounds.  Each symbol represents a polarization vector, 
indicated by $P$. 
The box highlights the values obtained with ``Wannier anions'' (see text).
(c) Sketch of the zigzag interface between two compounds (A$_a$B$_a$ in red and A$_b$B$_b$ in blue). 
}}
\end{center}
\end{figure}
According to the modern theory of polarization~\cite{Vanderbilt1993}, and
as further elucidated in Refs.~\cite{Stengel2009,Bristowe2011,Stengel2011c},
the net charge at the interface between two compounds
can be exactly given by invoking only bulk properties of the parent materials.
In particular, one needs to consider the lattice (Fig.~\ref{honeycomb})
of allowed values for the so-called \emph{formal} polarization, ${\bf P}$, of
either crystalline constituent.
Then, based on the interface orientation and termination~\cite{Stengel2011c}, one
can readily deduce the interface net charge.
An intuitive way to do this is by representing the electronic structure of the material in terms
of localized \emph{Wannier functions}~\cite{Marzari1997,Stengel2006}.
For polarization purposes, these effectively map the continuous charge density distribution of the
periodic crystal into a discrete set of classical point charges, 
located at the Wannier centers~\footnote{The Wannier calculation was performed with 
the Wannier90~\cite{Mostofi2008} code and the density functional theory (DFT) code 
{\sc Siesta}~\cite{Ordejon1996,Soler2002}.} shown in Fig.~\ref{honeycomb}(a).
Then the formal polarization of the crystal is simply given by the total dipole moment of the 
ensemble of these charges and the ion cores divided by the unit cell volume.
Of course, there are infinite possible choices for the primitive basis of atoms
and Wannier orbitals, hence the multivaluedness of the formal ${\bf P}$
[see Fig.~\ref{honeycomb}(b)].

A particularly convenient approach is that of combining the valence Wannier
functions with their nearest neighbor ion core, 
obtaining a negatively charged object that we call ``Wannier anion''~\cite{Stengel2011c}.
This way, we  obtain polarization lattices that include only a subset of the original
points [larger symbols in Fig.~\ref{honeycomb}(b)].
More specifically, each of these restricted points simply corresponds to the
dipole moment of a system of two integer point charges $Q=\pm Ne$, placed at
two arbitrarily chosen lattice sites, where $N$ is the ``formal valence'' of
the cation in each compound.
Note that, due to the three-fold symmetry of these structures,
both the Wannier anion and the cation carry zero dipole moment;
hence, the formal polarization only contains the point-charge
contribution~\cite{Stengel2011c}. This has some analogies with the
LaAlO$_3$/SrTiO$_3$ case, where it was shown that covalency
effects are also irrelevant to charge-counting purposes~\cite{Stengel2011c,Bristowe2009}.

Each arbitrary choice of atomic basis, in turn, yields a well-defined interface
termination once the unit cell is periodically repeated to generate a semi-infinite
domain.
For the \emph{zigzag} interfaces considered in this work,
the appropriate choice of the bulk unit cell
(i.e. one of the many choices that tile the whole system without leaving
any unpaired ion at the interface) is that shown by shaded regions in Fig.~\ref{honeycomb}(c).
The dipole moment along $y$ of such a unit cell is given by $d=-Qa/(2\sqrt{3})$,
yielding a formal polarization of $P_{y} = -Q/3a$. 
The interface theorem relates differences in formal polarization to the net
bound charge at the interface, and we have
\begin{equation}
\label{eqsigma}
\sigma_{\rm bound} = P_y^{(b)} - P_y^{(a)} = -\frac{\Delta Q}{3a}.
\end{equation}
This implicitly assumes that both parent compounds are
\emph{hexagonal} and have the same equilibrium lattice parameter. 
In practice there is always a small lattice mismatch, and to 
realize a pseudomorphic interface one or both constituents need to be strained.
Then, Eq.~(\ref{eqsigma}) must be corrected by piezoelectric contributions on $P_y$.

\begin{figure}[t]
\begin{center}
\includegraphics[
width=0.44\textwidth]{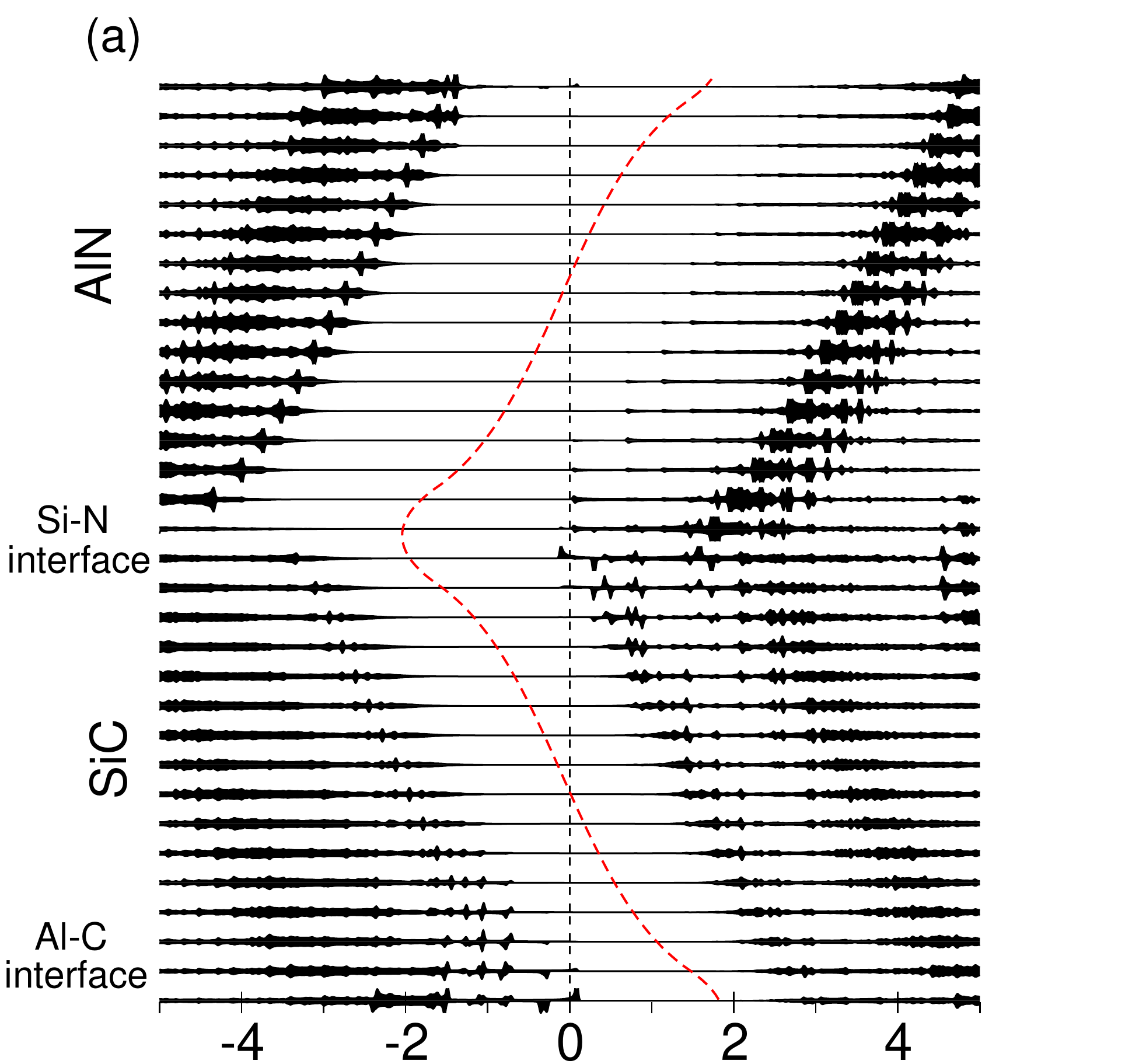}
\includegraphics[
width=0.42\textwidth]{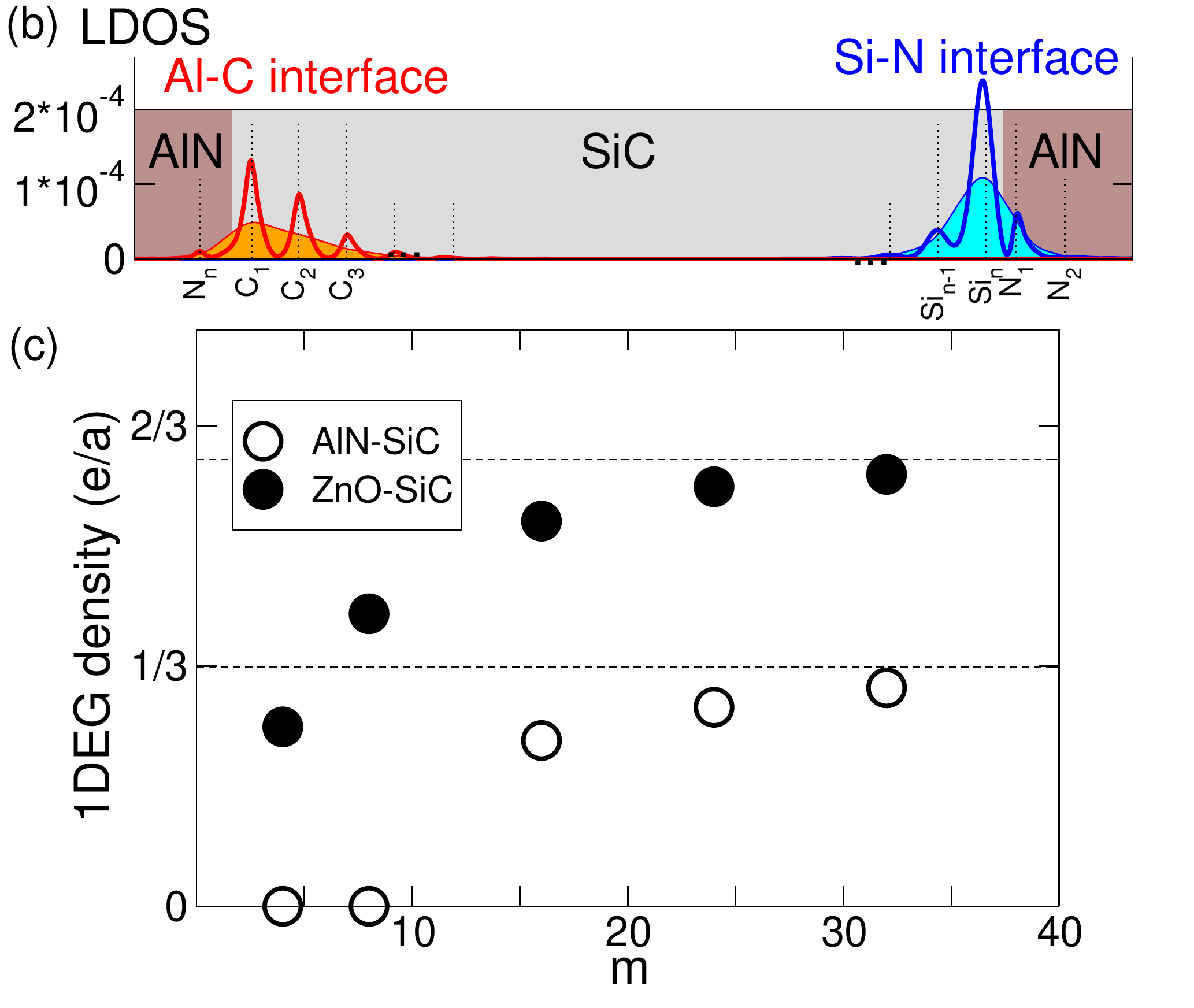}
\caption{\label{1DEG}{
(Color online) 
(a) Spin-resolved layer-by-layer density of states of the AlN-SiC ($m$=16) superlattice.
The top half corresponds to the AlN slab, and the bottom half to the SiC slab, 
with the interface terminations indicated.
Dashed (red) line schematizes the electrostatic potential, V(y), 
for an infinite array of alternating +/- cylindrical charges.
(b) The spatial (plane averaged) charge distribution computed from the local 
density of states (LDOS) 
is shown in the lower panel (shaded curves are macroscopically averaged LDOS). 
(c) The density of free electrons (or holes) calculated as a function of 
slab width ($m$) for the two superlattice systems studied (AlN-SiC and ZnO-SiC). 
Theoretical values corrected by piezoelectric effects are marked by dashed lines.
}}
\end{center}
\end{figure}

%
The electrostatic potential produced by this interfacial net charge diverges in the limit
of increasingly wide domains and needs to be somehow neutralized.
To investigate possible compensation mechanism, we use explicit first-principles 
calculations of 2D superlattices, which are constructed by periodically repeating 
two compositionally distinct nanoribbons along the stacking direction ($y$), 
so that the interface direction ($x$) follows the zigzag edge of the hexagonal lattice 
[see Fig.~\ref{honeycomb}(c)], consistent with the above discussion of the bulk polarity.
Having a relatively low ($<$7\%) lattice mismatch, we chose SiC, AlN and ZnO as
representative compounds for groups IV-IV, III-V and II-VI respectively.
The periodicity of the superlattices is controlled by varying the nanoribbon width, $m$ (given by 
the number of AB zigzag chains).
Only stoichiometric compounds are considered, which means that two chemically distinct 
interfaces are present in our superlattices: one of type A$_a$B$_a$A$_b$B$_b$
(A$_a$: cation of material $a$, B$_b$: anion of material $b$ etc.) and one of type
A$_b$B$_b$A$_a$B$_a$.
Along the normal direction to the sheet we include a vacuum layer of 20\AA{}, 
sufficiently thick to obtain converged results 
(no significant differences are observed with a vacuum layer of up to 200\AA).
The calculations are performed using the spin-polarized Perdew-Burke-Ernzerhof (PBE) exchange correlation 
functional~\cite{Perdew1996} as implemented in the {\sc Siesta} code~\cite{Ordejon1996,Soler2002}. 
Norm conserving pseudopotentials~\cite{Troullier1991} and a double-$\zeta$ polarized basis set are
used to represent the valence electrons~\cite{Anglada2002}. 
Atomic forces are relaxed to less than 20 meV/\AA .


%
Figure \ref{1DEG}(a) shows the spin-resolved layer-by-layer density of states (DOS) of the 
AlN-SiC ($m$=16) superlattice (III-V / IV-IV). 
The presence of a macroscopic sawtooth-like potential, corresponding to electric fields of
opposite polarity in the two material regions, is clearly visible from the plot.
These fields are generated by the net interfacial charges discussed above.
Under this potential the valence band maximum (VBM) at the $p$-type (Al-C) interface 
eventually becomes higher in energy than the conduction band minimum (CBM) at the $n$-type (Si-N) 
interface, and a charge transfer becomes favorable in energy.
The resulting accumulation of free carriers along the interfaces originate the
aforementioned 1DEG/1DHG whose spatial localization 
can be appreciated from their calculated planar-averaged densities, plotted in Fig.~\ref{1DEG}(b).
%
The 1DEG/1DHG, in turn, partially neutralize the net charge at either interface, 
reducing the potential offset.
This feedback mechanism results in an effective \emph{pinning} of the $p$-VBM and the $n$-CBM,
which remain close in energy regardless of the value of $m$ after breakdown occurs.
Thus, many aspects of the present electronic mechanism resemble those already reported for 
oxide superlattices~\cite{Bristowe2009}. 
However, we stress that the electrostatics of this system is different, as evidenced by 
the macroscopic potential profile, V(y), followed by the DOS, 
which is analogous to that of an infinite array of alternating positive and negative linear
charge densities [sketched in Fig.~\ref{1DEG}(a)].

At smaller ribbon widths (not shown) the associated potential drop is smaller than the electronic 
band gap of either participating compound, and the system can remain insulating (i.e. no ``Zener
breakdown'' occurs). 
%
In the large $m$ limit, on the other hand, the 1DEG/1DHG densities tend to the ``ideal'' value, which 
corresponds to perfectly neutralizing the discontinuity in 
the normal component of the bulk ${\bf P}$, according to Eq.~\ref{eqsigma}. 
This dependence of the charge transfer with ribbon width is shown in
Fig.~\ref{1DEG}(c) for both AlN-SiC (III-V / IV-IV) and ZnO-SiC (II-VI / IV-IV) interfaces.
This nicely shows a cross-over scenario from a short-period uncompensated regime, 
to a longer-period regime where a Zener-like breakdown mechanism occurs, similar to the
case of a LaAlO$_3$-SrTiO$_3$ superlattice.
The precise details of the cross-over depends on the band gaps and alignments between the two materials, 
which might not be properly given by DFT calculations, although the physical mechanism remains valid.
%
Note that the asymptotic limits (dashed lines in Fig.~\ref{1DEG}(c)), in fact, do not correspond
exactly to the ideal rational values ($1/3$ and $2/3$) of Eq.~\ref{eqsigma}. This is due to piezoelectric
effects, which again are accounted for by macroscopic bulk properties of the participating materials.
In particular, the lattice mismatch between SiC and ZnO, forces a compression in the ZnO ribbon 
(and a minor expansion in SiC)
resulting in a shift of its formal polarization lattice along the axis normal to the interface, 
hence decreasing $P_{\text{SiC}}-P_{\text{ZnO}}$. 

\begin{figure}[b]
\begin{center}
\includegraphics[width=3.45in]{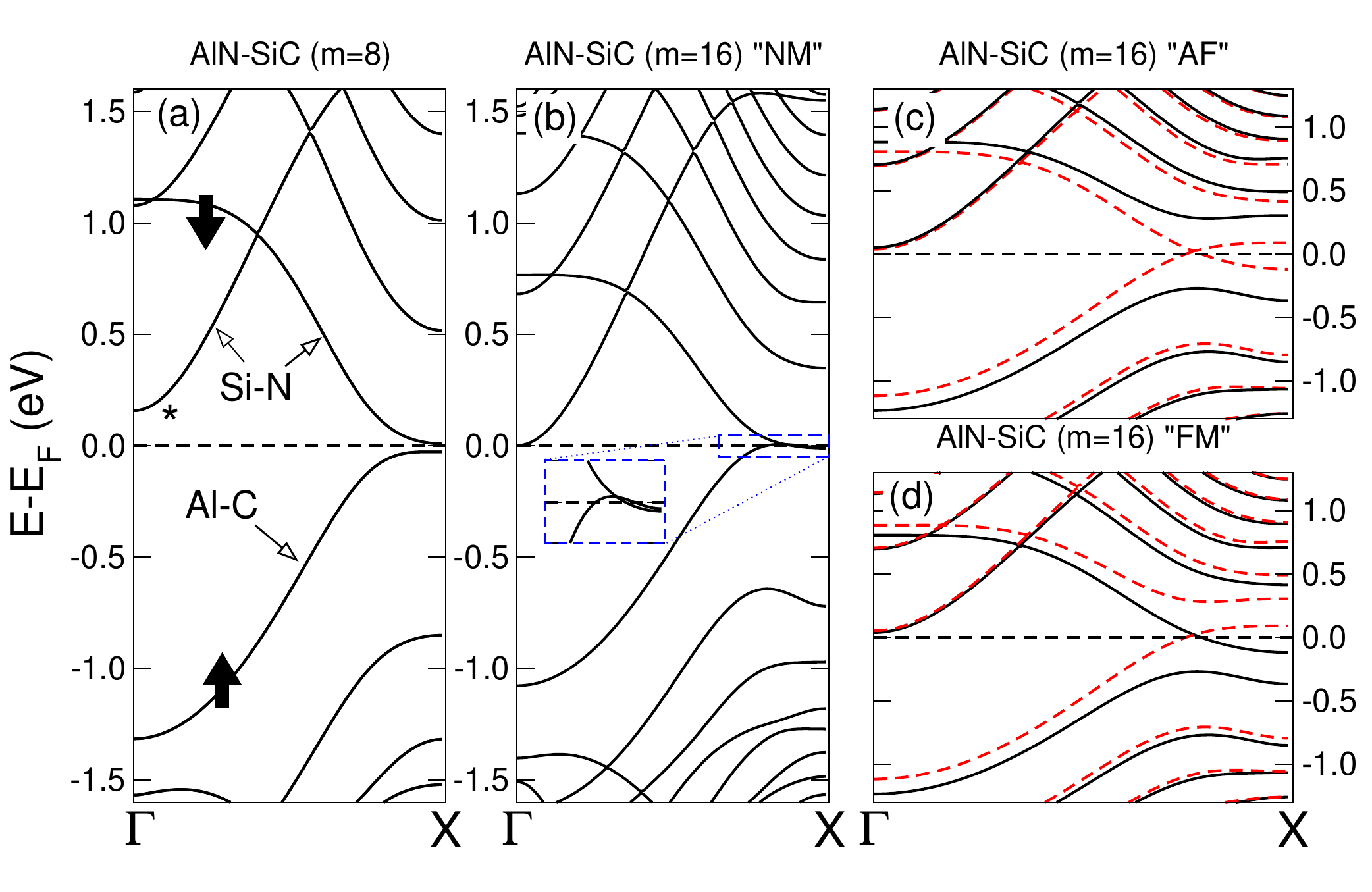}
\caption{\label{bands}{(Color online)
The band structure of the AlN-SiC superlattices for $m=8$ (a), and $m=16$ with paramagnetic-NM (b), 
antiferromagnetic-AF  (c) and ferromagnetic-FM (d) solutions. 
Solid (black) and dashed (red) lines correspond to bands with different spin components. 
The inset in (b) shows the Fermi level crossing of bands at the zone boundary.
}}
\end{center}
\end{figure}

We find that the 1D gases of carriers at the interfaces of these insulating materials 
are fully spin-polarized. To further investigate magnetic effects,
we plot in Fig.~\ref{bands} the band structure near the Fermi level, $E_f$, for a AlN-SiC superlattice
with thickness that corresponds to the uncompensated regime (a), and a partially compensated 
superlattice with larger thickness (panels (b)-(d) in the figure).
There are three dispersive $\pi$ bands close to $E_f$ that are involved in the compensation mechanism,
one occupied and mostly localized at the Al-C interface, and two empty and mostly localized at the Si-N interface. 
%
%
The flat bands at $X$ for the non-magnetic solutions Fig.~\ref{bands}(a) and \ref{bands}(b) 
very much resemble the edge states in graphene nanoribbons described with a simple tight-binding model 
for the $\pi$ electrons~\cite{Fujita1996}. The corresponding high-density of electronic states at $E_f$
suggests magnetic ordering due to a Stoner instability. 

Indeed, two spin-polarized phases with ferromagnetic orderings {\it along the interface} 
are lower in energy than the paramagnetic solution, and correspond to the transfer of charge (electrons) 
with well defined spin from one interface ($p$-VBM) to the other ($n$-CBM):
when the spin is preserved in the transfer, as in Fig.~\ref{bands}(c), 
the electronic magnetization at both interfaces is antiferromagnetically aligned (AF), 
whereas a spin-flip of the transferred charge gives the ferromagnetic (FM) solution 
(panel d). 
The two solutions, AF and FM {\it between} interfaces, are almost degenerate in energy, 
because the 1DEG and the 1DHG are decoupled by the wide insulating ribbon.  
Notice that the whole system is half-metallic for the AF solution, 
with full spin polarization of the conducting electrons.
%
%
%
%
%

It is generally agreed that a one-dimensional metal is unstable with respect to a symmetry lowering 
modulation of the charge~\cite{Peierls1955}. The coupling between lattice vibrations and electrons
near the Fermi level drives Peierls distortions that open a gap. 
Such deformation could also be a way to resolve the instability due to the large density of 
states at the Fermi level shown in Fig.~\ref{bands}(b).
To address the stability of the 1DEG against Peierls distortions, calculations using 
zone-folding techniques were performed for the AlN-SiC ($m$=16) superlattice, including up to three or 
four unit cells along the $x$ axis (in the limit of large $m$ a supercell with three unit cells should 
be enough to accomodate symmetry-lowering deformations capable of opening a gap in the right region 
of the Brillouin zone). 
We did not detect any trace of bond-length alternations in our tests, 
all of which remained in a metallic state. This is in agreement with previous studies 
that showed that Peierls distortions can open a gap in similar edge bands 
only for narrow ($<$6\AA) graphene~\cite{Pisani2007} and graphane nanoribbons~\cite{Tozzini2010}.

%
%

%
%

In addition to competing structural/electronic ground states, another factor that could thwart experimental 
realization of the 1DEG/1DHG predicted here are defects. 
In particular, these polar structures may, in certain growth conditions, find more energetically 
favorable compensation mechanisms than electronic transfer. 
For example an anionic vacancy could donate electrons to the $n$-type interface, or combinations 
of anionic vacancies and cationic vacancies at each interface could provide the necessary charge 
compensation without free electrons. Similar mechanisms have been proposed at the LaAlO$_3$-SrTiO$_3$ 
interface and in ferroelectric thin films~\cite{Bristowe2011b,Fong2006}.
Substitutional defects and charge transfers between edge states in isostructural C/BN superlattices 
have been also discussed~\cite{Pruneda2012}. If the growth conditions can be controlled such that 
these point defects are minimized it may be possible to utilize the systems here for various 
applications such as spintronics, sensors and electronics.  

In conclusion, we have shown that polar compensation mechanisms can give rise to fully spin-polarized 
one-dimensional electron and hole gases at the interface between two coplanar insulating bidimensional 
materials, in close analogy with the 3D case of LaAlO$_3$/SrTiO$_3$(001) interface.
The net charge at the interface can be determined solely from bulk properties (the polarization) 
of the parents materials, as required by the interface theorem~\cite{Vanderbilt1993}. 
The polarization is given by a simple charge counting procedure, 
irrespective of covalency/ionicity, that results in three distint sets of 
allowed polarization lattices, for the II-VI, III-V and IV-IV compounds, 
and to the formation of new electronic states (1DEGs) at the interface 
between two materials of different classes.

\begin{acknowledgments}
We acknowledge
I. Souza for valuable discussions,
the support of EPSRC, MCINN (FIS2009-12721-C04-01 and CSD2007-00041) and computing 
resources of CamGRID and Darwin at Cambridge, the 
Spanish Supercomputer Network and HPC Europa.
PBL acknowledges DOE support under FWP 70069.
\end{acknowledgments}


\end{document}